\documentclass[a4paper,doc,natbib]{apa6}

\usepackage[english]{babel}
\usepackage[utf8x]{inputenc}
\usepackage{amsmath}
\usepackage{graphicx}
\usepackage[colorinlistoftodos,prependcaption]{todonotes}
\usepackage{url}
\usepackage{booktabs}
\usepackage{caption}
\usepackage{framed}
\usepackage{subcaption}
\usepackage{verbatim}
\usepackage{epigraph}
\usepackage{soul}
\usepackage{nicefrac}

\newcommand{\given}{\, | \,}

\newcommand{\si}{\sigma}
\newcommand{\Ga}{\Gamma}

\title{J.\ B.\ S.\ Haldane's Rule of Succession}
\shorttitle{J.\ B.\ S.\ Haldane's Rule}
\author{Eric--Jan Wagenmakers$^1$, Sandy Zabell$^2$, and Quentin F. Gronau$^3$}
\affiliation{$^1$ University of Amsterdam\\
$^2$ Northwestern University\\
$^3$ The University of Newcastle, Australia
~\\Correspondence concerning this manuscript should be addressed to:
  ~\\E.-J. Wagenmakers ~\\University of Amsterdam, Department of Psychology
  ~\\Nieuwe Achtergracht 129B ~\\1018VZ Amsterdam, The Netherlands ~\\E--mail may be sent to EJ.Wagenmakers@gmail.com.}

\abstract{After Bayes, the oldest Bayesian account of enumerative induction is given by Laplace's so-called \textit{rule of succession}: if all $n$ observed instances of a phenomenon to date exhibit a given character, the probability that the next instance of that phenomenon will also exhibit the character is $\frac{n+1}{n+2}$. Laplace's rule however has the apparently counterintuitive mathematical consequence that the corresponding  ``universal generalization'' (every future observation of this type will also exhibit that character) has zero probability. In 1932, the British scientist J.\ B.\ S.\ Haldane proposed an alternative rule giving a universal generalization the positive probability $\frac{n+1}{n+2} \times \frac{n+3}{n+2}$. A year later Harold Jeffreys proposed essentially the same rule in the case of a finite population. A related variant rule results in a predictive probability  of $\frac{n+1}{n+2} \times \frac{n+4}{n+3}$. These arguably elegant adjustments of the original Laplacean form have the advantage that they give predictions better aligned with intuition and common sense. In this paper we discuss J.\ B.\ S.\ Haldane's rule and its variants, placing them in their historical context, and relating them to subsequent philosophical discussions.}

\begin{document}
\maketitle
The problem of enumerative induction has fascinated philosophers from antiquity to modern times. One touchstone is how such studies deal with  a general law or so-called \emph{universal generalization} (UG), in which all instances or observations of a phenomenon share a particular characteristic. For concreteness, consider the Goldbach conjecture, which states that every even integer greater than 2 can be expressed as the sum of two primes (for example, $8 = 5 + 3$; $50 = 47 + 3 = 37 + 13 = 31 + 19$). The Goldbach conjecture has not yet been proven, but it has been verified for all consecutive even numbers up to $4 \cdot 10^{18}$, which means that we know of $2 \cdot 10^{18} - 1$ confirming instances and zero disconfirming instances \citep{OliveiraEtAl2014}. Each confirming instance presumably makes the Goldbach conjecture more plausible. But how large is the increase in plausibility? And how likely is it that some specific, as-yet-unchecked number (e.g., ${10^{10}}^{100}$; \citealp[p. 9172]{Paseau2021}) is consistent with the Goldbach conjecture?

The problem of enumerative induction was addressed by \citet{Laplace17741986} who assigned the unknown proportion $\theta$ of confirmatory instances a uniform beta$(1,1)$ prior distribution.\footnote{There are some subtleties here centering around whether one is sampling from a finite or so-called hypothetical infinite population.  These are briefly discussed at the end.  For the moment we will implicitly assume the latter.}   Observing $n$ confirmatory instances results in a $\theta \sim \text{beta}(n+1,1)$ posterior distribution on $\theta$;   using the notation $z \in \text{UG}$ to indicate the next observation is confirmatory, the probability $P(z \in \text{UG} \given n)$ that the next instance $z$ will also be confirmatory is then given by \emph{Laplace's rule of succession} (\citealp[p.\ 19]{Laplace18141902}; see generally \citealp{Zabell1989}):  
\begin{equation}\label{eq:Laplace}
\begin{split}
    P(z \in \text{UG} \mid n) &= \frac{n+1}{n+2}\\
    & = 1 - \frac{1}{(n+2)}.
\end{split}
\end{equation}

Eq.~\eqref{eq:Laplace} shows that as $n$ becomes large, the probability the next instance is confirmatory approaches 1. However, this standard Bayesian analysis implies from the outset that the probability of the UG is zero. To see this, suppose we want to predict whether an entire sequence $z'$ of new instances will be confirmatory. The probability all $z'$ instances are confirmatory is easily seen to be $(n+1)/(n+z'+1)$ (see, e.\ g., \citealp[Appendix II]{Jeffreys1973}). As the sequence of $z'$ instances under consideration increases, the probability that \emph{all} instances in the sequence are confirmatory decreases to zero and the probability of finding a violation increases to one. Another way of viewing the same problem is to consider a sequence of $z'=n+1$ new instances; the probability that all instances in this sequence are confirmatory is only $\nicefrac{1}{2}$. ``This shows that the analysis of sampling procedure given so far is quite inadequate to account for the high probability that we often attach to a general law'' (\citealp[p. 53]{Jeffreys1973}; see also \citealp[pp. 127-128]{Jeffreys1961} and the corresponding fragments in earlier editions \citealp[pp. 111-113]{Jeffreys1948} and \citealp[pp. 106-109]{Jeffreys1939}).

Many scholars---including Laplace himself---recognized early on that Laplace's rule did not apply to settings in which additional background knowledge suggests the universal generalization could be true (see \citealp[pp. 296-7]{Zabell1989}, and references therein). One prominent example is the critical discussion of Augustus De Morgan, who promoted Laplace's work on probability theory in England: 

\begin{quotation}
If as before, the first $m$ Xs observed have all been Ys, and we ask what probability thence, and thence only, arises that the next $n$ Xs examined shall all be Ys, the answer is that the odds in favour of it are $m+1$ to $n$, and against it $n$ to $m+1$. No induction then, however extensive, can by itself, afford much probability to a universal conclusion, if the number of instances to be examined be very great compared with those which have been examined. If 100 instances have been examined, and 1000 remain, it is 1000 to 101 against all the thousand being as the hundred.

This result is at variance with all our notions; and yet it is demonstrably as rational as any other result of the theory. The truth is, that our notions are not wholly formed on what I have called the \emph{pure induction}. In this it is supposed that we know no reason to judge, except the mere mode of occurrence of the induced instances. Accordingly, the probabilities shown by the above rules are merely \emph{minima}, which may be augmented by other sources of knowledge. For instance, the strong belief, founded upon the most extensive previous induction, that phenomena are regulated by uniform laws, makes the first instance \emph{of a new case}, by itself, furnish as strong a presumption as many instances would do, independently of such belief and reason for it. \citep[pp. 214-215]{deMorgan18472003}
\end{quotation}

It was obvious therefore, Harold Jeffreys argued, that the relatively low confidence expressed by Laplace's rule that the next instances are confirmatory ``does not correspond to my state of belief or anybody else's'' \citep[p. 107]{Jeffreys1939}. It was evident that in order to explain the discrepancy between Laplace's rule and our intuition, the UG needed to be taken seriously in the sense of being given some separate prior mass (e.g., \citealp{WrinchJeffreys1921}, as inspired by \citealp{Broad1918}; see also \citealp[pp. 29-31]{Jeffreys1931}).

\section{Haldane's Solution}
Even though many scholars believed that Laplace's rule of succession often violates common sense, it appears that it was only in 1932 that J.\ B.\ S.\ Haldane first proposed a specific, concrete alternative to Laplace's rule to address this issue.\footnote{Remarkably, this is the same paper in which Haldane computes the first Bayes factor. \citet{EtzWagenmakers2017} detail Haldane's contribution and describe Haldane's mixture prior, but they stop short of commenting on the specific rule of succession that Haldane presents.} The simplest way to understand Haldane's alternative is as follows. We adopt the usual default assumptions (these can easily be relaxed): (1) before seeing any instances, the probability that the UG is correct equals $\nicefrac{1}{2}$, that is, $P(\text{UG}) = P(\neg \text{UG}) = \nicefrac{1}{2}$; (2) under the hypothesis that the UG is incorrect, every proportion $\theta$ of confirmatory instances is deemed equally likely \emph{a priori}: $\theta \given \neg \text{UG} \sim \text{beta}(1,1)$. 

In this framework, on the hypothesis that the UG is false, we obtain Laplace's rule of succession;  on the hypothesis that it is true, the probability of a confirmatory instance equals 1. The missing piece of the puzzle is the posterior probability that the UG is correct. When the UG is correct, $\theta = 1$. This is a risky hypothesis that makes a single prediction. In contrast, under the hypothesis that the UG is incorrect the proportion $\theta$ is assigned a uniform distribution, whose lack of commitment is reflected in predictions that are uniform across all possible $n+1$ outcomes. Consequently, the observation of $n$ confirmatory instances yields evidence in favor of the UG; specifically, the \emph{Bayes factor} (\citealp{EtzWagenmakers2017,Jeffreys1939,KassRaftery1995}) is $n+1$ (e.g., \citealp[p. 59]{Haldane1932}; \citealp[p. 55]{Jeffreys1973}) and this raises the probability for the UG from $\nicefrac{1}{2}$ to $(n+1)/(n+2)$. Interestingly, the result is numerically identical to Laplace's rule of succession, but it now refers to the posterior probability that the UG is correct and  \emph{all} unseen instances will prove to be confirmatory. (This numerical concordance explains what De Morgan has in mind for the last sentence of the quotation from him given earlier.)

Our end goal was to quantify our confidence that a specific unseen instance (e.g., for the Goldbach conjecture, a large number such as ${10^{10}}^{100}$) will be confirmatory when checked. However, we are uncertain about whether or not the UG is correct; in addition, when the UG is incorrect we are uncertain about the true proportion of numbers that are confirmatory. To obtain the desired result we model-average by applying the law of total probability:
\begin{equation}\label{eq:Haldane}
\begin{split}
P(z \in \text{UG} \mid n)& = \overbrace{\frac{n+1}{n+2}}^{P(\text{UG} \given n)}  \,\, \times \overbrace{1}^{P(z \in \text{UG} \given \text{UG}, n)}\,\,\, + \,\,\, \overbrace{\frac{1}{n+2}}^{P(\neg \text{UG} \given n)} \,\, \times \overbrace{\frac{n+1}{n+2}}^{P(z \in \text{UG} \given \neg \text{UG}, n)}\\
& = \,\, \frac{n+1}{n+2} \,\,\, \times \,\, \frac{n+3}{n+2}\\
& = \left[1 - \frac{1}{n+2}\right] \,\, \times \,\, \left[1 + \frac{1}{n+2}\right]\\
& = 1 - \frac{1}{(n+2)^2},
\end{split}
\end{equation}
an expression we term \emph{Haldane's rule of succession}. The second line of the equation highlights that including the hypothesis that the UG is correct yields a particularly simple result: Laplace's rule of succession needs to be adjusted by a multiplicative factor of $(n+3)/(n+2)$. The third line of the equation shows that the Laplacean first factor and the Haldanean second factor are symmetric about 1. The fourth line of the equation underscores that as $n$ grows, Haldane's rule of succession is associated with an increase in confidence that is more pronounced than it is for Laplace's rule of succession. Thus, the probability of finding an exception ``is clearly of the order $n^{-2}$, rather than $n^{-1}$. This seems to be a more reasonable estimate of the validity of an induction than that generally given'' \citep[p. 59]{Haldane1932}. 

The exercise can of course be repeated using different priors. For examples, we can use any prior odds $d = P(\text{UG})/P(\neg \text{UG})$ differing from 1; doing so changes the adjustment term in the second line from the equation from 
\[
\frac{n+3}{n+2} \qquad \text{to} \qquad
\frac{d(n+2)+1}{d(n+1)+1}.
\]
 
Alternatively, one may generalize the assumption concerning the shape of the beta distribution;  for example, instead of $\theta \given \neg \text{UG} \sim \text{beta}(1,1)$ we consider $\theta \given \neg \text{UG} \sim \text{beta}(\alpha,1)$ for $\alpha>1$. Larger values of $\alpha$ mean that $\neg \text{UG}$ is increasingly similar to UG in terms of its predictions for future instances. According to the Laplacean analysis that mathematically implies that the UG is false, we have that $P(z \in \text{UG} \given \neg \text{UG},n,\alpha) = \frac{n+\alpha}{n+\alpha+1}$. The Bayes factor in favor of the UG is $\frac{n}{\alpha} + 1$. A derivation similar to Eq.~\ref{eq:Haldane} then yields the generalized expression:
\begin{equation*}
\begin{split}
    P(z \in \text{UG} \mid n, \alpha) &= \frac{n+\alpha}{n+\alpha+1} \times \frac{n+2\alpha+1}{n+2\alpha}\\
    &= \left[1 - \frac{1}{n+\alpha+1}\right] \,\, \times \,\, \left[1 + \frac{1}{n+2\alpha}\right],
\end{split}
\end{equation*}
where the first factor represents the Laplacean answer and the second factor gives the Haldanean adjustment. When $\alpha=1$, we naturally recover Eq.~\ref{eq:Haldane}; when $\alpha=2$, we find that $P(z \in \text{UG} \mid n, \alpha=2) = \frac{n+2}{n+3} \times \frac{n+5}{n+4}$; and when  $\alpha=3$, we find that $P(z \in \text{UG} \mid n, \alpha=3) = \frac{n+3}{n+4} \times \frac{n+7}{n+6}$.  

Irrespective of the details of the prior assumptions, an increase in the number of confirmatory instances always has two effects: (1) the UG becomes increasingly plausible; (2) the posterior distribution of $\theta$ under $\neg \text{UG}$ is shifted toward 1. For example, in the case of the Goldbach conjecture the latter effect means that even if the Goldbach conjecture were shown to be false (which is highly unlikely, given the evidence at hand), the abundance of confirmatory instances would still make it highly likely that ${10^{10}}^{100}$ is the sum of two primes. 

The difference between Laplace's rule and Haldane's rule becomes more pronounced as the number of to-be-predicted instances increases. Consider the probability that an entire sequence of $z'$ new instances are all confirmatory. As indicated above, Laplace's rule gives $P(z' \in \text{UG} \given n) = (n+1)/(n+z'+1)$. Under Haldane's setup, however, we obtain
\begin{equation*}
\begin{split}
P(z' \in \text{UG} \mid n)& = \overbrace{\frac{n+1}{n+2}}^{P(\text{UG} \given n)}  \,\, \times \overbrace{1}^{P(z' \in \text{UG} \given \text{UG}, n)}\,\,\, + \,\,\, \overbrace{\frac{1}{n+2}}^{P(\neg \text{UG} \given n)} \,\, \times \overbrace{\frac{n+1}{n+z'+1}}^{P(z' \in \text{UG} \given \neg \text{UG}, n)}\\
& = \,\, \frac{n+1}{n+z'+1} \,\,\, \times \,\,\, \frac{n+z'+2}{n+2},
\end{split}
\end{equation*}
where the second factor represents the Haldanean adjustment. For instance, when $n=10$ and $z'=11$ the Laplacean probability that $z' \in \text{UG} \given n$ equals only $\nicefrac{1}{2}$, in violation of common sense (e.g., \citealp[pp. 127-128]{Jeffreys1961} and references above); in contrast, the Haldanean probability equals $\nicefrac{1}{2} \times \nicefrac{23}{12} = \nicefrac{23}{24} \approx 0.96$. In general, when $z' = n+1$ the Laplacean analysis gives $P(z' \in \text{UG} \given n) = \nicefrac{1}{2}$ whereas the Haldanean analysis gives $P(z' \in \text{UG} \given n) = \nicefrac{1}{2} + \nicefrac{1}{2} \cdot \frac{n+1}{n+2}$, an upward adjustment equal to half of the Laplacean probability that the single next observation is confirmatory. When we consider $\theta \given \neg \text{UG} \sim \text{beta}(\alpha,1)$ for $\alpha>1$, the generalized expression equals
\begin{equation*}
P(z' \in \text{UG} \mid n, \alpha) = \,\, \frac{n+\alpha}{n+\alpha+z'} \,\,\, \times \,\,\, \frac{n+2\alpha+z'}{n+2\alpha}.
\end{equation*}

In our notation, \citet[p. 59]{Haldane1932} gives the probability of the next instance being an exception as:
\begin{equation*}
    P(z \notin \text{UG} \mid n) = \frac{P(\neg \text{UG}) \int_0^1 \theta (1-\theta)^k \,\text{d}\theta}{P(\text{UG}) + P(\neg \text{UG}) \int_0^1 (1-\theta)^k \,\text{d}\theta},
\end{equation*}
which he mistakenly evaluates to $1/[(n+2)(n \cdot P(\text{UG}) +1)]$. The correct solution, however, is 
\[
\frac{1-P(\text{UG})}{(n+2)(n \cdot P(\text{UG}) +1)},
\]
 matching the results reported above  (the term in the numerator is $1 - P(UG)$, not 1).\footnote{As is easily verified using the standard formula for the beta integral when the parameters are integers. The fact that Haldane's solution is incorrect is apparent by considering the special case $P(\text{UG}) = 1$, when the probability of a disconfirming instance should be zero. This is presumably due to a typographical error.} In a response to Haldane one year later, \citet[pp. 85-86]{Jeffreys1933prior} proposed a similar rule for the finite population case, assuming that $P(\theta = 0 \given \text{UG}) = P(\theta = 1 \given \text{UG}) = P(\neg \text{UG}) = \nicefrac{1}{3}$ (cf. \citealp[pp. 128-132]{Jeffreys1961}).  

We can adjust the Haldane-Jeffreys setup in a number of simple ways. For instance, instead of assigning prior mass of $\nicefrac{1}{3}$ to each of the three hypotheses, we can instead assign prior mass $\nicefrac{1}{2}$ to $\neg UG$ and divide the remaining $\nicefrac{1}{2}$ equally over $\theta=1$ and $\theta=0$, which can be viewed as opposite versions of a universal generalization. Thus we have $P(\neg \text{UG}) = P(\text{UG}) = \nicefrac{1}{2}$ and $P(\theta = 0 \given \text{UG}) = P(\theta = 1 \given \text{UG}) = \nicefrac{1}{2}$. When $n=0$ there are no data and the Bayes factor equals 1; when $n \geq 1$ the Bayes factor equals $(n+1)/2$. We obtain:
\begin{equation}\label{eq:EJ}
\begin{split}
P(z \in \text{UG} \mid n)& = \overbrace{\frac{\tfrac{1}{2}(n+1)}{1+\tfrac{1}{2}(n+1)}}^{P(\text{UG} \given n)}  \,\, \times \overbrace{1}^{P(z \in \text{UG} \given \text{UG}, n)}\,\,\, + \,\,\, \overbrace{\frac{1}{1+\tfrac{1}{2}(n+1)}}^{P(\neg \text{UG} \given n)} \,\, \times \overbrace{\frac{n+1}{n+2}}^{P(z \in \text{UG} \given \neg \text{UG}, n)}\\
& = \,\, \frac{n+1}{n+2} \,\,\, \times \,\, \frac{n+4}{n+3}\\
& = 1 - \frac{1}{\tfrac{1}{2}(n+2)(n+3)},
\end{split}
\end{equation}
for $n \geq 1$. The additional vagueness with respect to the nature of the general law has resulted in an adjustment factor to Laplace's rule of succession that is less pronounced than before (as $(n+4)/(n+3) < (n+3)/(n+2)$ for $n \geq 0$).\footnote{Essentially the cost of a single observation.} As before, this approach may be generalized by replacing $\theta \given \neg \text{UG} \sim \text{beta}(1,1)$ by $\theta \given \neg \text{UG} \sim \text{beta}(\alpha,1)$ for $\alpha>1$. The Bayes factor in favor of the UG then equals $(\frac{n}{\alpha}+1)/2$, and we obtain the following expression for the probability that the next instance is confirmatory:
\begin{equation*}
\begin{split}
    P(z \in \text{UG} \mid n, \alpha) &= \frac{n+\alpha}{n+\alpha+1} \times \frac{n+3\alpha+1}{n+3\alpha}\\
    &= \left[1 - \frac{1}{n+\alpha+1}\right] \,\, \times \,\, \left[1 + \frac{1}{n+3\alpha}\right],
\end{split}
\end{equation*}
When $\alpha=2$, we find that $P(z \in \text{UG} \mid n, \alpha=2) = \frac{n+2}{n+3} \times \frac{n+7}{n+6}$; and when  $\alpha=3$, we find that $P(z \in \text{UG} \mid n, \alpha=3) = \frac{n+3}{n+4} \times \frac{n+10}{n+9}$. 


\section{Later Developments}

In a sequence of $n$ Bernoulli trials with \emph{unknown} success probability $\theta$, the standard Bayesian analysis expresses the probability of $S_n = k$ successes as
\[
P(S_n = k) = \int_0^1 \dbinom{n}{k} \theta^k (1-\theta)^{n-k} \, d\mu(\theta), \quad 0 \leq k \leq n,
\]
where $d\mu$ is the initial (or ``prior'') probability  for $\theta$. Such an expression in principle leaves open the status of both $\theta$ and $\mu$. The parameter $\theta$ is commonly viewed as a \emph{physical} entity, an aspect of the world, either a propensity, frequency in a finite population, or the limit of an infinite sequence, while the prior could be either physical or \emph{subjective}, in the latter case summarizing our beliefs.  But a purely subjective analysis is also possible (avoiding any mention of physical probabilities), and is worth briefly discussing here as an entry into a subsequent and substantial philosophical literature into which Haldane's rule can be attractively placed.

\subsection{The de Finetti Representation}

A sequence of random variables  $X_1, X_2, \dots, X_n$ is said to be \emph{exchangeable} if its distribution is invariant under permutation:  if $\si$ is a permutation of the integers $1, 2, \dots, n$, then $X_{\si(1)}, X_{\si(2)}, \dots, X_{\si(n)}$ has the same distribution as $X_1, X_2, \dots, X_n$;  and an infinite sequence $X_1, X_2, \dots, X_n, \dots$ is exchangeable if every finite subsequence of it exchangeable.  Assume for simplicity for the moment that the $X_k$ are 0-1 values, and that $S_n = X_1 + \dots + X_n$ records the number of 1s.  The \text{de Finetti representation theorem}  tells us that for such an infinite sequence:

\begin{enumerate}
\item The sample frequency $S_n/n$ converges almost surely:  For some random variable $Z$, 
\[
P\left(\lim_{n \to \infty} S_n/n  \to Z \right) = 1.
\]
\item If $\mu$ is the distribution of $Z$, so that $P(Z \in A) = \mu(A)$, then for all $n \geq 1$,
\[
P(S_n = k) = \int_0^1 \dbinom{n}{k} \theta^k (1-\theta)^{n-k} \, d\mu(\theta), \quad 0 \leq k \leq n.
\]
\end{enumerate}
This remarkable result has several important consequences from the subjectivist perspective:

\begin{itemize}
\item It gives a subjective explanation for the existence of infinite limiting frequencies, which are seen to be a consequence of the infinite exchangeability of the sequence $X_1, X_2, \dots$.  (Strictly speaking, only the stationarity of the sequence is being used here, in conjunction with the ergodic theorem.)
\item It gives a subjectivist explanation for the role of parameters in statistical inference. (The discussion here has focused on the binomial case, but a similar analysis can be advanced for a number of other common statistical families;  see \citealp{DiaconisFreedman1980}; \citealp[Chapter 7]{Jeffrey1992}; \citealp{Zabell2011}).
\item Under mild conditions the posterior distribution of $\theta$ can be shown to center more and more tightly around the observed sample frequency as the size of the sample $n \to \infty$;  thus de Finetti's theorem gives a solution to Hume's problem of induction (why should the future resemble the past?);  see Zabell (1988 and 1989) for discussion.\nocite{Zabell1988,Zabell1989}
\end{itemize}

This leaves open the question of the choice of the prior $\text{d}\mu(\theta)$. Consider the case of $t \geq 3$ possible outcomes rather than just two (say $c_1, \dots, c_t$), and let $n_1, n_2, \dots, n_t$ represent the frequencies with which these outcomes occur in a sample of size $n$ (so that necessarily $n = n_1 + \dots + n_t$).   In this case the representation takes the form of a mixture of multinomial probabilities
\[
\dbinom{n}{n_1 n_2 \dots n_t} \theta_1^{n_1}\theta_2^{n_2} \dots \theta_t^{n_t} 
\]
over the probability simplex 
\[
\Delta_t := \{(\theta_1, \theta_2, \dots, \theta_t) : \sum_i \theta_i = 1 \; \text{and} \;  \theta_i \geq 0, \; 1 \leq i \leq t \}.
\]


Let $\Ga(x)$ denote the gamma function defined on the positive axis $x > 0$ and  let $k_i > 0, \; 1 \leq i \leq t,$ be a sequence of $t$ positive constants.  The conjugate Dirichlet prior 
\[
d\mu(\theta_1, \dots, \theta_t) := \frac{\Ga(\sum_{i=1}^t k_i)}{\prod_{i=1}^t \Ga(k_i)} \theta_1^{k_1 - 1} \dots \theta_t^{k_t - 1} d\theta_1 \dots d\theta_{t-1}
\]
has the attractive property that (letting $k := k_1 + \dots + k_t)$ the predictive probabilities for an outcome of the $j$-th type occurring on the next trial is 
\[
P(X_{n+1} = c_j \mid n_1, \dots, n_t) = \frac{n_j + k_j}{n + k};
\] 
Such priors can already be found in Laplace, but because this is a continuous distribution on the simplex $\Delta_t $, the universal generalizations $UG_i$ for $1 \leq i \leq t$, that is, the infinite sequences $(c_i, c_i, c_i, \dots, )$, all have probability 0. 

\section{Predictive Probabilities Without Priors}

The tradition of using a prior on a simplex to derive predictive probabilities goes back to Bayes and Laplace, but during the twentieth century a completely different approach was advanced by the philosophers William Ernest Johnson (1858--1931) and Rudolf Carnap (1891--1970).  
Suppose as before one has a sequence $X_1, X_2, \dots $ taking values $c_1, c_2, \dots, c_t$, multinomial probabilities $\theta = (\theta_1, \dots, \theta_t) \in \Delta_t$,  a Dirichlet prior on $\Delta_t$ with parameters $k_1, k_2, \dots, k_t$ , and a sample of size $n$ with $n_j$ outcomes of type $c_j$ .  Then the sequence $X_1, X_2, \dots$ has the following properties:
\begin{enumerate}
\item It is exchangeable: for every $n$, and sequence of possible outcomes $e_1, \dots, e_n$, the joint probabilities $P(X_1 = e_1, \dots, X_n = e_n)$ are the same for all permutations of $e_1, \dots, e_n$;
\item  The cylinder sets have positive probability:  $P(X_1 = e_1, \dots, X_n = e_n) > 0$;
\item The probability of seeing type $j$ on the next trial, given a sample of size $n$  with frequency counts $n_1, n_2, \dots, n_t$, only depends on the sample size $n$ and the frequency counts $n_i$.
\end{enumerate}
This last property reflects the fact that the predictive probabilities are given by the formula
\begin{equation}
P(X_{n+1} = c_j \mid n_1, \dots, n_t) = \frac{n_j + k_j}{n + k},
\end{equation}
where $k = k_1 + \dots k_t$. 

In the 1920s Johnson was able to show that these three properties essentially characterize the class of Dirichlet priors:  if  $X_1, X_2, \dots$ satisfies the above three properties, and $t \geq 3$ (there are at least three types), then either the sequence is independent (so that the future does not depend on the past) or there exist positive constants $k_i > 0$ such that property (1) holds for all $n$ and all frequency counts $n_1, \dots, n_t$.\footnote{Johnson suffered a stroke in 1927 before he was able to publish his result, and his paper appeared posthumously in \emph{Mind} in 1932, edited for publication by R.\ B.\ Braithwaite;  see \citet{Johnson1932}, \citet{Zabell1982}.}  The celebrated de Finetti representation theorem tells us that every infinitely exchangeable sequence can be represented as a mixture of multinomial distributions;  and (4) in turn  tells us  that the mixing measure is the Dirichlet distribution with parameters $k_1, \dots, k_t$.   The third property is known as \textit{Johnson's sufficientness postulate} \citep{Zabell1982}, and Johnson's theorem tells us that it singles out the Dirichlet priors.  This beautiful result thus gives a subjectivist justification for using Dirichlet priors for reasons other than mathematical convenience and the conjugate prior property, and does not invoke the dubious principle of insufficient reason.

In the years leading up to 1952 the famous philosopher Rudolf Carnap independently followed Johnson's steps, coining the predictive probabilities as \textit{the continuum of inductive methods} (\citealp{Carnap1952}; \citealp{Zabell1997}). Because the Dirichlet prior assigns zero probability to the vertices of the simplex $\Delta_t$ (it is, after all, continuous), this means that Carnap's system does as well, and a number of critics pointed to this property as a drawback of the system.  (All this in apparent ignorance of Haldane's paper, although of course all this took place in the setting of qualitatively characterizing the predictive probabilities rather than starting from the setting of a prior on probabilities.)  In the 1960s and 1970s however, several mathematically trained philosophers were able to show that if one weakened the sufficientness postulate so that the predictive probabilities depended not only on $n$ and $n_i$, but also the \textit{number} of types $T$ observed so far, then this characterized the prior as being in general a mixture of:  (a) point masses on the vertices, (b) beta distributions on the edges, three-dimensional Dirichlet distributions on the faces, and so on, up to the full $t$-dimensional Dirichlet distribution on $\Delta_t$.  

Thus, as long as one sees only one type ($T = 1$), then some positive probability is assigned to universal generalizations;  but as soon as $T=2$, universal generalizations are naturally assigned zero probability, but some probability is still assigned to the possibility that only the two types observed so far will continue to be observed.  
References to this philosophical literature include \citet{Johnson1932}, Carnap (1950 and 1952)\nocite{Carnap1950,Carnap1952}, \citet{Hintikka1966}, \citet{HintikkaNiiniluoto1976}, \citet{Niiniluoto1981}, \citet{Niiniluoto2011}, and \citet{Kuipers1978}. \citet{Zabell1996} shows that the narrow issue of permitting the existence of universal generalizations (without the full apparatus of Hintikka, Niiniluoto, and Kuipers) was possible even at the time Carnap wrote:  the part of the sufficientness postulate that rules out universal generalizations is easily identified and---once removed---Johnson's proof can be relatively easily extend to show that universal generalizations are now included in the range of possibilities.  For a detailed discussion of the history of the Johnson and Carnap results, and the amusing way in which Carnap's work almost exactly paralleled that of Johnson's earlier results and the path he took, see \citet{Zabell2011}.

\section{The Role of the Infinite}

Infinite sequences and infinite limiting frequencies are useful fictions so long as the presence of the infinite does not qualitatively impact the analysis.  This is the case here.

\subsection{The Role of the Integral Representation}

The de Finetti representation of a finite exchangeable sequence as an integral mixture can fail if the sequence is not infinitely exchangeable (that is, is not the initial segment of an infinitely exchangeable sequence).  This can certainly happen:  for example, if one draws a random sample without replacement from an urn containing $R$ red balls and $B$ black, and $N = R + B$, then it is not hard to see that the resulting probability on $N$-long sequences is exchangeable but cannot be further extended in exchangeable fashion.  (Intuitively, this is because after $N$ draws the urn is empty, but it is also a simple analytic fact easily established.  Consider for example the trivial case when $R = B = 1$.)

This would appear to be a limitation in using the representation theorem, but there is in fact an attractive finite version of the theorem which clarifies the situation.  Diaconis and Freedman (1980) show that if an exchangeable 0-1 sequence $X_1, \dots, X_k$ can be extended to an exchangeable sequence $X_1, \dots, X_n$ ($n \geq k$), then there exists a mixture of Bernoulli trials such that the variation distance between the distribution of $X_1, \dots, X_k$ and the mixture restricted to the first $k$ outcomes is bounded by $4k/n$.  More generally, if the exchangeable sequence $X_1, \dots, X_k$ takes its values in a finite set $S$ of $t \geq 2$ elements and has distribution $P$ on $S^k$,  but can be extended to an exchangeable sequence $X_1, \dots, X_n (n \geq k$), then there is a multinomial mixture such that the variation distance between $P$ and $Q^k$, the multinomial mixture restricted to $S^k$, satisfies the simple bound
\[
|| P - Q^k|| \leq 2tk/n.
\]
In other words, the more the original sequence can be extended, the better the possible approximate de Finetti representation that can be found.    So for much the same reasons that we use the Poisson distribution (as a large sample approximation to an underlying binomial reality), it is reasonable to work with the de Finetti representation.  (And in any case, the Johnson-Carnap and Hintikka-Niiniluoto theorems also have finite versions.)

\subsection{Just What is a Universal Generalization?}

The Wrinch-Jeffreys-Haldane approach seems to say that a universal generalization corresponds to putting positive probability mass on certain points in the simplex (or, more generally, in the Hintikka approach, to sub-simplexes).  But there are some subtleties here.  What does it mean to put positive mass on $\theta =1$?  If we think of $\theta$ as an infinite limiting frequency, then the universal generalization might fail;  after all, the sequence
\[
0100100000111111111111111111
\]
 (all 1's after the first ten outcomes) has a limiting frequency of 1s of 1.  Worse, there might an infinite number of exceptions:  if in 
 \[
 0110101000101000101000100000 ... \]
 a ``1'' represent the prime occurrences (and 0 composite), then the prime number theorem tells us that limiting frequency of 1s in this sequence is 0.  What one really means by an infinite generalization is that in sequence space the specific sequence 1111111111111111111 is assigned positive probability.  

The predictive probability approach championed in various ways by Pearson, Johnson de Finetti, and later Carnap avoids these issues by replacing limiting results by appropriate finite versions. 

\section{Nothing New Under the Sun}

Haldane's approach seems so natural that it seems surprising that prior to  Haldane (and to some extent Wrinch and Jeffreys) it had not been suggested.  So it should not be entirely surprising that at least the qualitative intuition behind it can indeed be found earlier---some 18 centuries earlier in fact!  

In his book \emph{On Medical Experience} the Greek physician Aelius Galenus (129--216 AD, ``Galen'') mounted a defense of the empiricist approach to medicine. One of his arguments in modern terminology goes as follows. He invites us to consider the first outcome of a trial (such as testing a new medicine). Then, Galen argues, there are four possibilities: in the future such an outcome will always occur, generally occur, be equally likely to occur or not, or rarely occur. Until we make further observations there is no way to know which of these will be the case.   Deductive reasoning cannot decide the issue, only empirical observation.  The first of the four is what we have been calling a universal generalization, and Galen is saying that as long as only one kind of outcome is seen (for example, the medicine is always effective), this provides evidence (qualitatively of course for Galen)  in support of the universal generalization that it will always work in the future.  It is only when other outcomes are observed (for example, the patient dies) that the universal generalization is rejected, and then further observation is needed to distinguish between the other three cases.  Here are Galen's actual words (translated from the Greek of course):

%
\begin{quotation}
For things which are visible fall into four classes:  one of them is always plain, another generally so, in the case of another lucidity and obscurity are equally balanced, and the fourth is rarely plain.  If then something visible to the eye is seen only once, this single observation will not suffice to indicate which of the four kinds of medical science it belongs to.  Since we do not know that it will appear on every future occasion as it has done on this occasion, how should we know that it is always thus? Therefore it is not possible for us after having seen a thing once to be able to foretell that what was seen on this occasion will often be seen, and that its opposite will only be seen rarely, just as it is not possible to know whether the reverse is the case.  And since this belongs to what cannot be recognized by a single observation, so in the case of both what is more frequent and what is rare it is impossible to know the thing after seeing it only once, and likewise it is not possible for that thing to be known whose nature consists (both) in its being and in its not being (amphidoxos). [As translated by \citealp[p. 112]{Walzer1944}; `amphidoxos' means `doubtful' in Greek]
\end{quotation}
 
It is interesting to note that Galen affords the case $\theta = \nicefrac{1}{2} $ a special status, as in the generalizations of Haldane's rule discussed earlier.  Another special case he does not consider is that of the \emph{hapax legomenon}, that is, the first outcome never recurs.  In Galen's medical setting this seems of no particular interest,  but curiously one can find this seriously considered in the early days of mathematical probability.  In his appendix to Bayes's famous essay, Richard Price also considers one's epistemic situation after the first observation, but from a very different perspective.  Price argues that after the first observation, the significance of what is seen is that 

\begin{quotation}
 Antecedently to all experience, it would be improbable as infinite to one, that any particular event, beforehand imagined, should follow the application of any one natural object to another; because it would be an equal chance for any one of an infinity of other events. 
\dots  The first experiment supposed to be ever made on any natural object would only inform us of one event that may follow a particular change in the circumstances of those objects; but it would not suggest to us any ideas of uniformity in nature, or give us the least reason to apprehend that it was, in that instance or in any other, regular rather than irregular in its operations. But if the same event has followed without interruption in any one or more subsequent experiments, then some degree of uniformity will be observed; reason will be given to expect the same success in further experiments, and the calculations directed by the solution of this problem may be made. (Price, 1763, appendix to \citealp[pp. 408--409]{Bayes1763})
 \end{quotation}
 
Thus, from Price's novel and interesting perspective, the first observation of some phenomenon merely tells us that it is possible, and it is only after the \emph{second} observation of it that one is in the standard framework where a rule of succession even makes sense.  It initially seems hard to make sense of this in the setting of iid repeated sampling, but Price's idea can indeed be made sense of in terms of a sampling from a distribution which has a continuous component (observations from this give rise to the hapax legomena, since the probability that two observations from a continuous distribution are equal is zero) and a finite discrete component.  There are interesting connections here with the sampling of species problem and exchangeable random partitions;  see Zabell (1992, 1997). \nocite{Zabell1992Predicting,Zabell1997}
 
 
\section{Conclusion}
Despite dissatisfaction with Laplace's rule of succession because of its inability to confirm universal generalizations, it was not until 1932 that J. B. S. Haldane (following in the steps of Wrinch and Jeffreys) first proposed and studied more general mixture priors, concluding the probability of finding an exception is of order $n^{-2}$ rather than $n^{-1}$. Unfortunately, Haldane's presentation contains an error and was not given in the form of Eq.~\ref{eq:Haldane}. A year later Jeffreys gave the correct rule, but for the finite sample case, possibly obscuring that what is at play is a simple adjustment to Laplace's original rule. The citation record suggests that as far as the rule of succession is concerned, Haldane's contribution has been entirely forgotten.\footnote{The contribution of Jeffreys is occasionally acknowledged (e.g., \citealp[p. 121]{DiaconisSkyrms2018}).} We hope this note will correct the historical record and generate some interest in the fact that Haldane's approach yields a simple adjustment of the standard Laplacean rule that anticipated the work of Hintikka and Niiniluoto several decades later, as well as providing a simple justification for a natural inductive intuition going back at least as far as Galen. 

\bibliography{referenties}
\end{document}